\documentclass[prl,twocolumn,showpacs]{revtex4}
\usepackage{amsmath}
\usepackage{amssymb}
\usepackage{graphicx}

\begin{document}

\title{Observable Phase Entanglement}
\author{K. W. Chan}
\email{kwchan1@pas.rochester.edu}
\author{J. H. Eberly}
\affiliation{{Center for Quantum Information and Department of Physics
and Astronomy,} \\ {University of Rochester, Rochester, NY 14627 USA}}

\date{\today}

\begin{abstract}
Quantum entanglement can manifest itself in the narrowing of wavepackets. We define the phenomenon of phase entanglement and describe its effect on the interpretation of spatial localization experiments. 
\end{abstract}


\maketitle

Substantial progress has been made in the last decade in the understanding of quantum entanglement in continuous spaces. For example, criteria are now available that determine the existence of entanglement, or not, of bipartite states \cite{Braunstein-Pati, Duan-etal, Simon, Manchini-etal}. It is only in continuous spaces that the degree of entanglement is not bounded, and it is here that the interest in high entanglement is focused. This is where the quantum-classical border domain \cite{Zurek} is entered by the Einstein-Podolsky-Rosen (EPR) thought experiment \cite{EPR}, and examination of high entanglement realms associated with EPR physics has begun \cite{Eberly-etal03}. A substantial theoretical clarification was made earlier by Reid and Drummond \cite{Reid-Drummond} in the context of two-mode photonic field amplitudes, which can formally play the roles of position and momentum variables. More recent theoretical studies have explored various physical realizations analogous to the EPR scenario \cite{Chan-etal02, Law-Eberly04, Chan-etal04, Fedorov-etal}. Experimental advances include early observations of small degrees of continuum entanglement \cite{Ou-etal92, Silberhorn-etal, Bowen-etal}. Recent measurements have been made with down-conversion photons in the high entanglement domain \cite{Howell-etal04}, where the EPR uncertainty relation deviates strongly from the Heisenberg relation, and massive-particle observations are being proposed \cite{Raizen}. 

While momentum and position offer equivalent bases in the usual mathematical Fourier sense for discussions of quantum entanglement in continuous spaces, they frequently correspond to experimentally quite different opportunities or challenges. In this Letter we show that the operational effect of this difference is to expose an unexpected feature that we believe has not previously been noted. This feature is phase entanglement.  We define it and show that it should be observable in a massive-particle EPR experiment. 

Surprisingly, one can find phase entanglement even in Gaussian states \cite{Englert-Wodk} that describe very uncomplicated physical situations. Perhaps equally surprising, the original EPR scenario of fragmentation or ``breakup" of two massive particles is one of these. Let us consider a breakup process in which the two particles, initially joined as a single unit, are decomposed into non-interacting fragments that fly apart. Molecular dissociation provides a natural physical example. We can analyze the dynamics of this simple two body problem by using  center of mass and relative coordinates, in terms of which the joint wave function at breakup is separable: $\Psi(x_1, x_2; t=0^+) = \phi_{cm}(x_1 + x_2) ~\phi_{rel}(x_1 - x_2)$, where we have taken equal masses for simplicity in defining coordinates. However, there is no guarantee that the separability extends to the individual particles, and generally they are in fact entangled and $\Psi(x_1, x_2; t)$ cannot be written in the product form $\Phi_1(x_1,t) ~\Phi_2(x_2,t)$. 

To be specific, suppose that the joint state initially takes the following form in position space
\begin{equation}
	\Psi(x_1, x_2; t=0^+) \sim
		e^{-\frac{(x_1+x_2)^2}{4b^2}} e^{-\frac{(x_1-x_2)^2}{4a^2}} ,
\label{eq:psi_x0}
\end{equation}
where $a>0$ and $b>0$. 
We will consistently omit irrelevant normalization factors.
The first factor, the $cm$ part of the wavefunction, is a free-particle Gaussian.  The choice of the exact form of the relative part only affects quantitative details but not the interesting qualitative features (see \cite{Fedorov-etal}), so we have taken it to be Gaussian as well. Note that the entanglement in Eq.~(\ref{eq:psi_x0}) is controlled by the packet width parameters $a$ and $b$.  The two particles are initially uncorrelated (the joint wave function factors) when the cross term $x_1 x_2$ is absent from the exponent, which (for Gaussians) is automatically true in the special case when $a=b$, so we assume $a \ne b$. The product of two Gaussian functions is also Gaussian, of course, and many analyses of Gaussian states are available, and a great deal is known about conditions on their entanglement \cite{Englert-Wodk, Giedke-etal01, Adesso-etal, Giedke-etal03}. We are raising a new point here.

After breakup the two particles are massive and free, so their positional wavepackets spread when they separate from each other. The nature of their free space evolution is well known. It will be useful to reproduce it here, approaching it first from the Fourier transform of Eq.~(\ref{eq:psi_x0}):
\begin{equation}
	\tilde{\Psi}(k_1, k_2; t=0^+) \sim
		e^{-\frac{a^2}{4}(k_1-k_2)^2} e^{-\frac{b^2}{4}(k_1+k_2)^2} ,
\label{eq:psi_k0}
\end{equation}
where $p_i = \hbar k_i$ is the momentum of particle $i$. It evolves with the free space Hamiltonian:
\begin{equation}
	H = \frac{p_1^2}{2m} + \frac{p_2^2}{2m} .
\end{equation}
Since the momenta $p_1$ and $p_2$ are quantum non-demolition (QND) operators
for the free space Hamiltonian, the time evolution only adds a phase to $\tilde{\Psi}$.
We have
\begin{equation}
	\tilde{\Psi}(k_1, k_2; t) \sim
	e^{-\frac{iQ^2}{2}\left(k_1^2+k_2^2\right)}
	e^{-\frac{a^2}{4}(k_1-k_2)^2} e^{-\frac{b^2}{4}(k_1+k_2)^2} ,
\label{eq:psi_kt}
\end{equation}
where $Q$ is the quantum diffusion length:
\begin{equation} \label{eq:QDef}
Q \equiv \sqrt{\frac{\hbar t}{m}}.
\end{equation}
By Fourier transform we quickly obtain the corresponding coordinate-space result:
\begin{equation} \label{eq:psi_xt}
	\Psi(x_1, x_2; t)
	\sim e^{-\frac{1}{4}~\frac{a^2(x_1 + x_2)^2 + b^2(x_1 - x_2)^2
		+ 2iQ(x_1^2 + x_2^2)} {a^2b^2 - Q^4 +iQ(a^2 + b^2)}} .
\end{equation}
We re-emphasize that these Gaussians are not those for harmonic oscillators but rather for free particles. The time dependence of the ever-growing quantum diffusion length $Q$ is important.

In order to comment on the quantum information contained in this time-dependent breakup state we will introduce the Schmidt decompositions \cite{SchmidtThm} of $\tilde\Psi(k_1,k_2;t)$ and $\Psi(x_1,x_2;t)$.  Since Eq.~(\ref{eq:psi_kt}) is similar to a two-mode squeezed state, the Schmidt basis here is also spanned by number states.  Indeed, using identities for Hermite functions, it is easy to find that the Schmidt modes $\tilde{\phi}_n(\kappa, \tau)$ are given by
\begin{equation}
	\tilde{\phi}_n(\kappa, \tau) = 
	\frac{1}{\pi^\frac{1}{4}}\sqrt{\frac{1}{2^n n!}}
	\ H_n(\kappa) e^{-\frac{\kappa^2}{2}} e^{-\frac{i}{2} \kappa^2 \tau} ,
\end{equation}
and they determine the characteristic single-sum Schmidt expansion:
\begin{equation}
	\tilde{\Psi}(k_1, k_2; t) = \alpha \sum_{n=0}^{\infty} \sqrt{\lambda_n}
		\ \tilde{\phi}_n\left(\alpha \, k_1, \frac{Q^2}{\alpha^2}\right)
		\, \tilde{\phi}_n\left(\alpha \, k_2, \frac{Q^2}{\alpha^2}\right) ,
\label{eq:Schmidt_k}
\end{equation}
where $\alpha = \sqrt{ab}$ and
\begin{equation}
	\lambda_n = \frac{4ab}{(a + b)^2}\Big( \frac{a - b}{a + b} \Big)^{2n} .
\end{equation}

The number of terms in the Schmidt sum is infinite, but since the $\lambda_n$'s are reduced density matrix eigenvalues we have $\sum_n\lambda_n =1$, and there is generally a small number that contribute significantly. This number is well estimated by what is reasonably called the Schmidt number $K$, given by 
\begin{equation} \label{eq:KDef}
	K \equiv \left(\sum_{n=0}^\infty \lambda_n^2 \right)^{-1}
	= \frac{1}{2}\left(\frac{a}{b}+\frac{b}{a}\right)
	= \cosh r ,
\end{equation}
where we defined $ e^r \equiv a/b $.
Similar expressions have been found useful in a study of transverse high entanglement in down conversion \cite{Law-Eberly04}. As we could have expected, $\tilde{\Psi}$ has the same Schmidt
form as a two-mode squeezed photon state, and the squeezing parameter is $r$. One has $1\le K<\infty$, so $K$ is a suitable choice for a quantitative measure of entanglement, and is also experimentally of direct relevance since it defines how many independent modes, i.e., Schmidt functions $\tilde{\phi}_n(\kappa, \tau)$, are important in any given study. 

Evaluations of entanglement can be associated with variances of Gaussian states in several ways \cite{Duan-etal, Simon, Manchini-etal, Adesso-etal}. Actual measurements of the degree of continuous entanglement are just beginning to be carried out in the domain of very high entanglement where one approaches the EPR limit of correlation. A high-entanglement measurement scenario for a pair of particles undergoing an EPR-type breakup has been described for the photon and recoiling atom in stimulated light scattering \cite{Chan-etal04} and for the signal and idler photons in parametric down conversion \cite{Law-Eberly04} for which the first related experimental results have recently appeared \cite{Howell-etal04}.  A theoretical proposal for a similar experimental determination, but for massive particles, was introduced in Ref.~\cite{Fedorov-etal}, and a cold optical lattice context for such an experiment has also been identified \cite{Raizen}. These situations are of wide interest because in the massive-particle case one will reproduce for the first time the exact EPR scenario. 

In all of these studies of EPR-like entanglement the roles played by conditional and unconditional variances are central \cite{Reid-Drummond}. 
In momentum space, the calculations are straightforward for conditional and unconditional variances for either of the two particles.  They correspond to measurements of dispersion in momentum that are either single-particle or coincidence measurements. Since the momentum space time dependence appears only in the phase of the two-particle wave function (\ref{eq:psi_kt}), the momentum variances  do not change as time evolves. We find for the single-particle dispersion
\begin{equation}
	\Delta^2 k_1^\text{single}
	\equiv \langle k_1^2 \rangle - \langle k_1 \rangle^2
	= \frac{a^2 + b^2}{4 a^2 b^2} = \frac{\cosh r}{2ab} .
\end{equation}
On the other hand, a coincidence measurement gives the conditional dispersion:
\begin{equation}
	\Delta^2 k_1^\text{coinc}
	\equiv \langle k_1^2 \rangle_{k_2} - \langle k_1 \rangle_{k_2}^2
	= \frac{1}{a^2+b^2} = \frac{1}{2\alpha\cosh r} ,
\end{equation}
where the expectation values are now calculated with respect to the
conditional probability $P(k_1 | k_2) = P(k_1, k_2)/P(k_2)$.
By combining these in the Fedorov ratio \cite{Fedorov-etal} we obtain
\begin{equation}
	R_p \equiv \frac{\Delta k_1^\text{single}}{\Delta k_1^\text{coinc}}
	= \cosh r = K .
\label{eq:Rp-K}
\end{equation}
This expression shows that the ratio of single to coincidence packet widths $R_p$, which is experimentally directly accessible, is a quantitative entanglement measure through its connection with the Schmidt number $K$.

Under some experimental conditions, it is easier to measure position variances. We evaluate them for our wave function and find:
\begin{eqnarray}
\hspace{-4mm} && \hspace{-1mm}
	\Delta^2 x_1^\text{single}(t)
	\equiv \langle x_1^2 \rangle - \langle x_1 \rangle^2 \nonumber \\
\hspace{-4mm} &=& \hspace{-1mm}
	\frac{a^2+b^2}{4a^2b^2}\left(a^2b^2 + Q^4\right)
		= \left(1+\frac{t^2}{t_0^2}\right)\frac{ab\cosh r}{2} ,
\end{eqnarray}
and
\begin{eqnarray}
\hspace{-5mm}	&& \hspace{-1mm}
	\Delta^2 x_1^\text{coinc}(x_2; t) \nonumber \\
\hspace{-5mm}	&\equiv& \hspace{-1mm}
	\langle x_1^2 \rangle_{x_2} - \langle x_1 \rangle_{x_2}^2 
	= \frac{\left(a^4 + Q^4\right)
		\left(b^4 + Q^4 \right)}
		{\left(a^2+b^2\right)\left(a^2b^2 + Q^4\right)}
		\nonumber \\
\hspace{-5mm}	&=& \hspace{-1mm}
	\frac{ab}{2\cosh r}
		\left(e^{2r}+\frac{t^2}{t_0^2}\right)
		\left(e^{-2r}+\frac{t^2}{t_0^2}\right)
		\left(1+\frac{t^2}{t_0^2}\right)^{-1} ,
\end{eqnarray}
where the expectation $\langle \cdots \rangle_{x_2}$ is with respect to the position conditional probability $P(x_1 | x_2)$, and we define $t_0 = mab/\hbar$.

\begin{figure}[!tb]
	\begin{center}
	\includegraphics[width=2.5in]{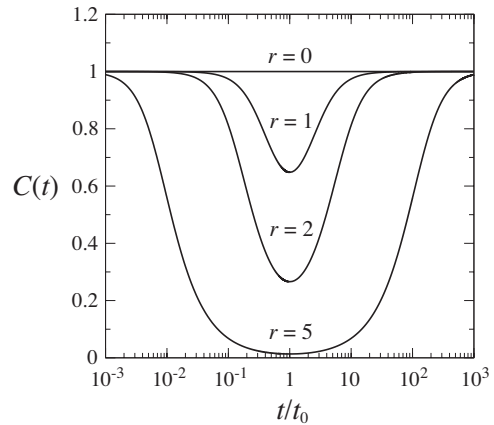}
	\caption{Plot of the function $C(t)$ in Eq.~(\ref{eq:CDef}) at
	$r=0, 1, 2 $ and $5$.  The time is in units of $t_0$.}
\label{fig:dx}
	\end{center}
\end{figure}

Entanglement is representation-independent (and time-independent) in such an interaction-free context as we are considering. Thus, given the simple Fourier-inverse character of momentum and position representations, it is surprising that $R_x = \Delta x_1^\text{single}(t)/\Delta x_1^\text{coinc}(x_2; t)$ does not exhibit the same properties as $R_p$. Not only that, the $x$-space width ratio is time-dependent: $R_x = K~C(t)$, where
\begin{equation}
	C(t) \equiv  \left(1+\frac{t^2}{t_0^2}\right) \Bigg/		\sqrt{\left(e^{2r}+\frac{t^2}{t_0^2}\right)
		\left(e^{-2r}+\frac{t^2}{t_0^2}\right)} .
\label{eq:CDef}
\end{equation}
A plot of the function $C(t)$ is shown in Fig.~\ref{fig:dx}.

The unexpected distinction between $R_p$ and $R_x$ reveals a novel element of free-particle Gaussian states (in contrast to the more commonly studied harmonic oscillator Gaussian states) that is open to experimental study, as we now explain.  If we inspect the probability density in position space (ignoring irrelevant normalization factors):
\begin{eqnarray} \label{eq:Px1x2}
\hspace{-1mm} && \hspace{-1mm}
	P(x_1, x_2; t) \equiv |\Psi(x_1, x_2; t)|^2 \nonumber \\
\hspace{-1mm} &\sim& \hspace{-1mm}
	\exp\left\{-\frac{1}{2}\frac{(a^2b^2 - Q^4)
		\left[a^2(x_1 + x_2)^2 + b^2(x_1 - x_2)^2\right]}
		{(a^2b^2 - Q^4)^2 + Q^4(a^2 + b^2)^2} \right\}
		\nonumber \\
\hspace{-1mm} & & \hspace{-1mm}
	\times \exp\left\{-\frac{1}{2}\frac{ 2Q^2\left(a^2 + b^2\right)
		\left(x_1^2 + x_2^2\right)}
		{(a^2b^2 - Q^4)^2 + Q^4(a^2 + b^2)^2} \right\} ,
\end{eqnarray}
we easily see that only the first factor carries the entangling $x_1x_2$ product term in the exponent. Therefore evidence of entanglement vanishes when the first exponent vanishes, i.e., when either (i) $a=b$, or (ii) $Q^2 = ab$.  The first case is trivial as remarked earlier. The second provides the new feature of Gaussian entanglement. It identifies the time $t=t_0$ at which entanglement information in $\Psi(x_1, x_2; t)$ in Eq.~(\ref{eq:psi_xt}) is completely transferred to the phase of the state.  At such a time none of the coherences in the two particle density matrix appear in the joint probability density (\ref{eq:Px1x2}). 

Moreover, this result makes it clear that not only at $t = t_0$ but at all times $t \ne t_0$ some portion of the entanglement is lost from the spatial joint probability, i.e., has become entanglement in positional phase alone. The function $C(t)$ is a natural measure of the extent to which entanglement information has been shifted into the phase of the two-particle state. Clearly we have the ordered inequality:
\begin{equation}
	K \equiv R_p \ge R_x \ge 1 ,
\end{equation}
and $C(t)$ shows that only two times, $t=0$ and $t\rightarrow\infty$, provide $R_x = R_p = K$.  The evolution of $R_x$ away from and then back to the value $R_p$ occurs for an interesting reason. It is the means by which the quantum state ensures a smooth transfer between two physically different conditions of breakup entanglement. 

These two conditions are easiest to see as distinct in the case of high entanglement, which occurs if $a \gg b$ or $b \gg a$. In either case the packet that initially has the much smaller width (either center of mass or relative) has to spread much faster than the other packet (Heisenberg uncertainty principle in action). Inevitably, what was the wider but slower-spreading packet becomes the narrower (Einstein delocalization and relocalization in action) and their roles in describing entanglement become reversed. Asymptotically their final ratio is just the reciprocal of their initial ratio, but as our expression (\ref{eq:KDef}) indicates, a reversal of the width ratio simply returns $K$ to the same value it had initially. We also can see that the trivial pure-phase condition $a=b$ also fits, because when the widths are initially equal they spread at the same rate and no Einstein localization ever takes place.

The Einstein conditional uncertainty and the unconditional Heisenberg uncertainty can now be compared for a general breakup scenario. The latter uncertainty is of course preserved throughout the coordinate-space shape-changing that occurs:
\begin{equation}
	\Delta^2 x_1^\text{single}(t) \cdot \Delta^2 k_1^\text{single}
	= \frac{\cosh^2 r}{4}\left(1+\frac{t^2}{t_0^2}\right)
	\ge \frac{1}{4} ,
\end{equation}
for all $t$ and $r$.
On the other hand, the conditional Einstein uncertainty relation reads
\begin{eqnarray}
\hspace{-7mm} && \hspace{-1mm}
	\Delta^2 x_1^\text{coinc}(x_2; t) \cdot \Delta^2 k_1^\text{coinc}(k_2)
	\nonumber \\
\hspace{-7mm} &=& \hspace{-1mm}
	\frac{1}{4\cosh^2 r}
	\left(e^{2r}+\frac{t^2}{t_0^2}\right)
	\left(e^{-2r}+\frac{t^2}{t_0^2}\right)
	\left(1+\frac{t^2}{t_0^2}\right)^{-1} ,
\end{eqnarray}
which has these limiting cases:
\begin{equation*}
\begin{array}{rll}
	\text{(i)} \ & t/t_0=0 , & \quad
	\Delta^2 x_1^\text{coinc} \Delta^2 k_1^\text{coinc}
	= \displaystyle \frac {1}{4\cosh^2 r} \le \frac{1}{4} ,
	\\[3mm]
	\text{(ii)} \ & t/t_0 = 1 , & \quad
	\Delta^2 x_1^\text{coinc} \Delta^2 k_1^\text{coinc}
	= \displaystyle \frac{1}{2} ,
	\\[3mm]
	\text{(iii)} \ & t/t_0 \gg e^r , & \quad
	\Delta^2 x_1^\text{coinc} \Delta^2 k_1^\text{coinc}
	\rightarrow \displaystyle \frac{t^2/t_0^2}{4\cosh^2 r} .
\end{array}
\end{equation*}

In summary we have examined both momentum and position variances arising in a  breakup experiment involving two massive particles, in the case that the particles exert little or no influence on each other following the breakup. This is a situation that can be realized in several ways, including molecular dissociation \cite{Fedorov-etal}.  Momentum and position are formally equivalent bases from which to examine entanglement, while they correspond to experimentally different scenarios. We showed that they also have a fundamental difference in phase entanglement which becomes evident in comparing the Fedorov width ratios for momentum and coordinate variances.  This can arise only in the massive free-particle coordinate-space case, exactly the situation of the original Einstein-Podolsky-Rosen thought experiment. Phase entanglement has not been discussed previously. We predict that it will be detectable in experiments designed to achieve high values of entanglement, of fundamental interest because of their nearness to the classical-quantum border \cite{Zurek}. 

Finally, it can be noted that there exist pure phase entangled states, e.g., 
\begin{equation} \label{eq:purephase}
\Psi \sim \exp\left\{-\mu^2 \left(x_1^2+x_2^2\right) + i \nu^2 x_1 x_2\right\}.
\end{equation}
This is very similar to an EPR state, except that $x_1$ is tightly
related to $k_2$ instead of $x_2$, as can be seen by taking a partial Fourier transform of $\Psi$ with respect to $x_2$.  For such a state the two-particle
probability shows no correlation between the two particles, and
entanglement tests based only on variances of position or momentum are unable to reveal its entangled nature. 

\bigskip
\noindent 
\textit{Acknowledgements}: We have benefitted from extended conversations and collaborations with M.V. Fedorov and C.K. Law about localization and entanglement and we thank K. Wodkiewicz for helpful remarks about Gaussian states. This work was supported by NSF grant PHY-0072359, a MURI grant administered as ARO-D DAAD19-99-1-0215, and the award to KWC of a Messersmith Fellowship.



\begin{thebibliography}{99}

\bibitem{Braunstein-Pati} See {\em Quantum Information with Continuous Variables}, edited by S. L. Braunstein and A. K. Pati (Kluwer, Dordrecht, 2003).

\bibitem{Duan-etal} L. M. Duan, G. Giedke, J. I. Cirac, and P. Zoller,
Phys. Rev. Lett. {\bf 84}, 2722 (2000).

\bibitem{Simon} R. Simon,
Phys. Rev. Lett. 84, 2726 (2000).

\bibitem{Manchini-etal} S. Manchini, V. Giovannetti, D. Vitali, and P. Tombesi,
Phys. Rev. Lett. \textbf{88}, 120401 (2002).
	
\bibitem{Zurek} W. H. Zurek, Phys. Today {\bf 44}, 36 (1991).

\bibitem{EPR} A. Einstein, B. Podolsky, and N. Rosen, Phys. Rev. {\bf 47},
777 (1935).

\bibitem{Eberly-etal03} J. H. Eberly, K. W. Chan, and C. K. Law,  
Phil. Trans. Roy. Soc. London A {\bf  361}, 1519 (2003).

\bibitem{Reid-Drummond}   M. D. Reid and P. D. Drummond, Phys. Rev. Lett. {\bf  60}, 2731 (1988); M. D. Reid, Phys. Rev. A {\bf  40}, 913 (1989).

\bibitem{Chan-etal02} K. W. Chan, C. K. Law, and J. H. Eberly, Phys. Rev. Lett.
{\bf 88}, 100402 (2002).

\bibitem{Law-Eberly04} C. K. Law and J. H. Eberly, Phys. Rev. Lett. {\bf 92}, 127903 (2004).

\bibitem{Chan-etal04} K. W. Chan, C. K. Law, M. V. Fedorov and J. H. Eberly, J. Mod. Optics {\bf XX}, yyyyyy (2004).

\bibitem{Fedorov-etal} M. V. Fedorov, M. A. Efremov, A. E. Kazakov,
K. W. Chan, C. K. Law, and J. H. Eberly, Phys. Rev. A {\bf 69}, xxxxx (2004).

\bibitem{Ou-etal92} Z. Y. Ou, S. F. Pereira, H. J. Kimble, and K. C. Peng,  Phys. Rev. Lett. {\bf 68}, 3663 (1992).

\bibitem{Silberhorn-etal} Ch. Silberhorn, P. K. Lam, O. Weiss, F. K\"onig, N. Korolkova, and G. Leuchs, Phys. Rev. Lett. {\bf 86}, 4267 (2001).

\bibitem{Bowen-etal} W. P. Bowen, R. Schnabel, P. K. Lam, and T. C. Ralph, Phys. Rev. A {\bf 69}, 012304 (2004).

\bibitem{Howell-etal04} J. Howell, R. S. Benninck, J. S. Bentley, and R. W. Boyd, Phys. Rev. Lett. {\bf 92}, aaaaa (2004).

\bibitem{Raizen} M. Raizen, private communication.

\bibitem{Englert-Wodk} For an extended overview of Gaussian states, see B.-G. Englert and K. Wodkiewicz, Int. J. Quant. Inf. {\bf 1}, 153-188 (2003).

\bibitem{Giedke-etal01} G. Giedke, B. Kraus, M. Lewenstein, and J. I. Cirac,
Phys. Rev. Lett. {\bf 87}, 167904 (2001).
	
\bibitem{Giedke-etal03} G. Giedke, M. M. Wolf, O. Kr\"uger R. F. Werner, and J. I. Cirac, Phys. Rev. Lett. {\bf 91}, 107901 (2003). 

\bibitem{Adesso-etal} G. Adesso, A. Serafini, and F. Illuminati,
Phys. Rev. Lett. {\bf 92}, 087901 (2004).

\bibitem{SchmidtThm} See M. A. Nielsen and I. L. Chuang, {\em Quantum Computation and Quantum Information} (Cambridge University Press, Cambridge, 2000), and A. Ekert and P. L. Knight, Am. J. Phys. {\bf 63}, 415 (1995). The continuum case is thoroughly detailed in S. Parker, S. Bose, and M. B. Plenio, Phys. Rev. A {\bf 61}, 032305 (2000).



\end{thebibliography}
\end{document}